\documentclass[sigconf]{acmart}

\copyrightyear{2026}
\acmYear{2026}
\setcopyright{cc}
\setcctype{by}
\acmConference[WWW '26]{Proceedings of the ACM Web Conference 2026}{April 13--17, 2026}{Dubai, United Arab Emirates}
\acmBooktitle{Proceedings of the ACM Web Conference 2026 (WWW '26), April 13--17, 2026, Dubai, United Arab Emirates}
\acmPrice{}
\acmDOI{10.1145/3774904.3792428}
\acmISBN{979-8-4007-2307-0/2026/04}

\newcommand{\ourname}{{ContRec}}

\usepackage{wrapfig}
\usepackage{caption}
\usepackage{enumitem}
\usepackage{threeparttable}
\usepackage{multirow}
\usepackage{algorithm}
\usepackage{algorithmic}
\usepackage{subfigure}
\usepackage[many]{tcolorbox}
\usepackage{caption}
\usepackage{balance}

\AtBeginDocument{%
  }

\begin{document}

\title{Diffusion Generative Recommendation with Continuous Tokens}

\author{Haohao Qu}
\email{haohao.qu@connect.polyu.hk}
\orcid{0000-0001-7129-8586}
\affiliation{%
  \institution{The Hong Kong Polytechnic University}
  \country{Hong Kong}
}

\author{Shanru Lin}
\email{lllam32316@gmail.com}
\affiliation{%
  \institution{City University of Hong Kong}
  \country{Hong Kong}
}

\author{Yujuan Ding}
\email{dingyujuan385@gmail.com}
\orcid{0000-0003-2945-1107}
\authornote{Corresponding author.}
\affiliation{%
  \institution{The Hong Kong Polytechnic University}
  \country{Hong Kong}
}

\author{Yiqi Wang}
\email{yiq@nudt.edu.cn}
\orcid{0000-0001-9594-1919}
\affiliation{%
  \institution{National University of Defense Technology}
  \country{China}
}

\author{Wenqi Fan}
\email{wenqifan03@gmail.com}
\orcid{0000-0002-4049-1233}
\affiliation{%
  \institution{The Hong Kong Polytechnic University}
  \country{Hong Kong}
}

\renewcommand{\shortauthors}{Haohao Qu, Shanru Lin, Yujuan Ding, Yiqi Wang, \& Wenqi Fan}
\begin{abstract}
Recent advances in generative artificial intelligence, particularly large language models (LLMs), have opened new opportunities for enhancing recommender systems (RecSys). Most existing LLM-based RecSys approaches operate in a discrete space, using vector-quantized tokenizers to align with the inherent discrete nature of language models.
However, these quantization methods often result in lossy tokenization and suboptimal learning, primarily due to inaccurate gradient propagation caused by the non-differentiable \texttt{argmin} operation in standard vector quantization.
Inspired by the emerging trend of embracing continuous tokens in language models, we propose \textbf{ContRec}, a novel framework that seamlessly integrates continuous tokens into LLM-based RecSys.
Specifically, \ourname{} consists of two key modules: a $\sigma$-VAE Tokenizer, which encodes users/items with continuous tokens; and a Dispersive Diffusion module, which captures implicit user preference.
The tokenizer is trained with a continuous Variational Auto-Encoder (VAE) objective, where three effective techniques are adopted to avoid representation collapse.
By conditioning on the previously generated tokens of the LLM backbone during user modeling, the Dispersive Diffusion module performs a conditional diffusion process with a novel Dispersive Loss, enabling high-quality user preference generation through next-token diffusion.
Finally, \ourname{} leverages both the textual reasoning output from the LLM and the latent representations produced by the diffusion model for Top-K item retrieval, thereby delivering comprehensive recommendation results.
Extensive experiments on four datasets demonstrate that \ourname{} consistently outperforms both traditional and state-of-the-art LLM-based recommender systems.
Our results highlight the potential of continuous tokenization and generative modeling for advancing the next generation of recommender systems.
\end{abstract}




\maketitle

\newcommand\webconfavailabilityurl{https://doi.org/10.5281/zenodo.17285302}
\ifdefempty{\webconfavailabilityurl}{}{
\begingroup\small\noindent\raggedright\textbf{Resource Availability:}\\
The source code of this paper has been made publicly available at \url{https://doi.org/10.5281/zenodo.17285302}.
\endgroup
}

\begin{figure}
    \centering
    \includegraphics[width=\linewidth]{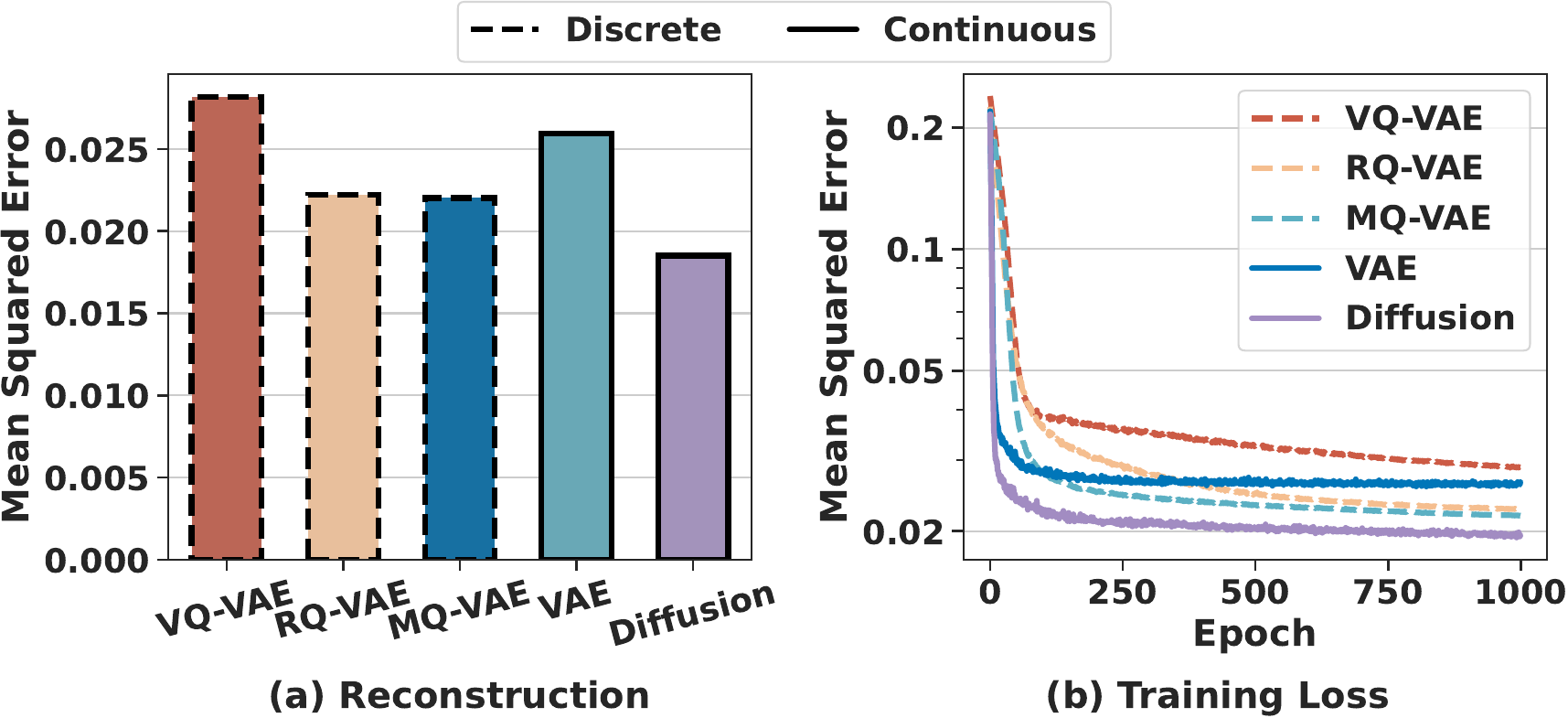}
    \captionsetup{font={small}}
        \vskip -0.1in
    \caption{Comparison of five representative deep generative models, namely VQ-VAE~\cite{van2017neural}, RQ-VAE~\cite{rajput2023recommender}, MQ-VAE~\cite{qu2024tokenrec}, VAE~\cite{kingma2014auto}, and Diffusion~\cite{ho2020denoising}, in reconstructing item embeddings on the Beauty dataset.
The standard diffusion model with continuous embeddings demonstrates superior reconstruction performance and loss convergence compared to the vanilla VAE and its discrete counterparts.}
    \label{fig:reconstruction}
    \vskip -0.1in
\end{figure}

\section{Introduction}

Web mining is a pivotal research area that focuses on extracting valuable knowledge and patterns from vast and heterogeneous web data.
Within this domain, recommender systems (RecSys) serve as crucial tools to tackle information overload and enrich user experiences on various web applications such as e-commerce, job search, and social media platforms~\citep{wang2020traffic,fan2019graph,fan2020graph}.
The emerging focus on Generative Artificial Intelligence (GenAI), especially Large Language Models (\textbf{LLMs}), has spurred a growing interest in \textbf{Generative Recommendation}, driven by the desire to leverage their exceptional generalization capabilities and adeptness at in-context learning~\citep{zhao2024recommender,fan2024graph}.
These distinct abilities empower LLM-based RecSys to generalize more effectively across new tasks and domains. 

Most existing LLM-based RecSys utilize LLM as the backbone, therefore inevitably inherit its core architecture design centered on discrete token representations~\citep{zhao2023survey,hirschberg2015advances}. 
For example, basic settings for item/user representation in RecSys include numeric indexing (e.g., "\emph{item\_2025}") and textual title indexing (e.g., "\emph{Apple iPhone 16, 1TB, White}")~\citep{hua2023index,geng2022recommendation,bao2023tallrec,wei2024llmrec}. 
However, these approaches become inefficient when dealing with large-scale real-world recommender systems, in which the sizes of user and item sets typically reach millions. 
To address the token set overload issue, Vector-Quantized Variational Autoencoder (VQ-VAE)~\citep{van2017neural, qu2024tokenrec} has been applied to construct a discrete vocabulary (also referred to as a "codebook", such as "$\langle a-3 \rangle \langle b-24 \rangle \langle c-58 \rangle \langle d-12 \rangle$") and train a neural tokenizer on users and items.
This strategy has been widely employed in a series of LLM-based RecSys, including TIGER~\citep{rajput2023recommender}, UTGRec~\citep{zheng2025universal}, and OneRec~\citep{deng2025onerec}, among others. 
These quantization-based solutions may be practical for encapsulating rich item-side information into a sequence of tokens that align with the discrete nature of language models, but they still encounter two critical limitations.
First, their reconstruction quality and generation performance are constrained by lossy tokenization, as some information will inevitably be compressed and omitted during the quantization  process~\cite{fan2024fluid,chen2025softvq,sun2024multimodal}.
Second, due to the non-differentiable \texttt{argmin} in typical vector quantization, a “straight-through” trick~\cite{bengio2013estimating} is normally used, directly copying the gradient from the decoder input to the encoder output.
Such a surrogate gradient can be inaccurate, potentially impeding both the achievement of high compression ratios and the effective learning of the latent space in quantization-based methods~\cite{huh2023straightening,fifty2024restructuring}.

Such limitations of discrete learning-based models have garnered attention in language tasks, especially for multimodal modeling (e.g., image, video, and protein)~\cite{wang2024diffusion,wang2025lavin}.
For example, \citet{fan2024fluid} reveal that quantization-based visual models exhibit poor performance improvements when scaling up model size, compared to models operating on continuous tokens.
Furthermore, there have been increasing efforts to introduce \emph{continuous tokens} for learning better representations~\cite{li2024autoregressive}. 
For example, \citet{chen2025softvq} introduce an efficient continuous image tokenizer for language models, which achieves a high compression ratio while enriching the semantic content of the latent space for image reconstruction. 
These advancements in language models inspire the exploration of continuous tokens as a new tokenization paradigm for LLM-based generative recommendation.
To achieve the goal, a few LLM-based recommendation methods have attempted to include continuous tokens. 
For example, CoLLM~\cite{zhang2023collm} and LlaRA~\cite{liao2024llara} propose directly projecting collaborative item representations learned from traditional recommendation methods into the token space of language models.
However, these methods only continualize the input part of the model by directly employing simple projection layers, e.g., a multi-layer perceptron (MLP), for user/item tokenization, while their output part inherits the LLM backbone's generator, which is learned for in-vocabulary discrete tokens. 
Such discrete-continuous gaps in one model might degrade its unification~\citep{sun2024multimodal} and result in suboptimal recommendation performance.

To enable the seamless incorporation of continuous tokens into LLM-based generative recommendation, we propose leveraging the powerful Diffusion Model to continualize user preference modeling and generation, alongside the tokenization of users and items.
It is widely recognized that user preference modeling is the core technical part leading to effective recommendation, while this task is highly challenging because user preferences are inherently complex and differences between individual users can be exceptionally subtle.
Diffusion Models are particularly promising for addressing this challenge, as they have demonstrated remarkable success in handling continuous tokens across various domains.
For instance, DEEM~\citep{luo2025deem} demonstrates that diffusion models can serve as the "eyes" of LLMs for image perception by encoding and decoding images as continuous tokens, while recent advances in protein language models demonstrate diffusion models' capability in constructing continuous protein structure embeddings~\citep{guo2024diffusion,wang2024dplm}. 
To validate this insight in recommendation scenarios, we conduct a preliminary item embedding reconstruction experiment on the Amazon-Beauty dataset.
As shown in Figure~\ref{fig:reconstruction}, our results are consistent with previous findings in the literature, showing that utilizing diffusion models with continuous representations delivers higher reconstruction accuracy and a smoother loss decline than the discrete counterparts.
Detailed information is provided in Section~\ref{sec:reconstruction} and Appendix~\ref{sec:app_result}.

In this paper, we present a novel LLM-based recommendation framework, \textbf{\ourname{}}, utilizing \textbf{Cont}inuous tokens to enhance the capabilities of LLM-based \textbf{Rec}Sys. 
With an LLM being the recommender backbone for user modeling with hybrid input, \ourname{} consists of two core continuous token learning components: a VAE module to learn continuous tokenization for users/items via the reconstruction of their associated features (e.g., IDs, texts, and images)
and a novel Disperse Diffusion module to learn continuous-distributed user preference representations based on the LLM-based user modeling.
Additionally, we design a hybrid retrieval mechanism that integrates the explicit LLM-output prediction text and the implicit diffusion-enhanced user preference representation to output a ranking list of items for precise recommendations.

Our major contributions are summarized as follows:
\vskip -0.1in
\begin{itemize}[leftmargin=*]
    \item We propose \ourname{}, a new framework that harnesses continuous tokens within LLM-based generative recommender systems, unlocking the advantages of non-quantized tokenization for users/items and enabling the generation of high-quality recommendations.
    By moving beyond the limitations of quantization, this solution paves the way for transformative advances in generative recommendation, offering a powerful and complementary perspective to existing discrete-based methods.

    \item We introduce two critical modules: a novel Dispersive Diffusion module and a robust $\sigma$-VAE tokenizer.
    The coefficient $\sigma$ is designed to safeguard against the collapse of continuous representations, ensuring rich and expressive latent spaces.
    The Dispersive Diffusion module is developed to learn implicit user preference representations through contrastive self-supervised learning without the need for explicit negative-positive pairs. 
    
    \item We conduct abundant experiments, including recommendation performance evaluation, ablation study, optimization robustness analysis, and hyperparameter tuning, to rigorously validate the effectiveness of the proposed method. The results demonstrate that \ourname{} outperforms the state-of-the-art baselines, notably surpassing the quantization-based method TIGER, with average improvements of 11.76\% on HR@10 and 10.11\% on NDCG@10 across four datasets.
    
\end{itemize}

\section{Preliminary}
\label{sec:pre}
\textbf{Generative Recommendations.}
This part recaps the basic notions of generative recommendations, as established by the pioneering LLM framework.
Consider $\mathcal{U}= \{u_1, u_2, ..., u_n\}$ and $\mathcal{V}= \{v_1, v_2, ..., v_m\}$ as the sets of users and items, respectively, where $n$ signifies the user count, and $m$ indicates the number of items in the dataset.
Among these, the item set associated with the historical interactions of user $u_i$ is symbolized by $\mathcal{V}_{(u_i)}$.
Generally speaking, recommender systems aim to grasp user preferences through analyzing interactions (e.g., clicks and reviews) between users $\mathcal{U}$ and items $\mathcal{V}$.
As a widely used solution, sequential recommendation techniques are developed to predict the future user preference from historical user-item interactions.
Correspondingly, we reformulate the sequential recommendation into a language model paradigm.
Given tokens $\mathcal{T}_i$ for each user $u_i$  and $\mathcal{T}_j$ for each item $v_j$, we can construct the integrated input for the LLM-based recommender with certain textual prompts $\mathcal{P}$ and obtain the output as:
\begin{align}
    \mathcal{Y}_i = \text{LLM}(\mathcal{P}(\mathcal{T}_i, \{\mathcal{T}_j|v_j \in \mathcal{V}_{(u_i)}\})).
    \label{eq:llm4rec}
\end{align}
The LLM output $\mathcal{Y}_i$ can take various forms, such as item titles, special tokens, or latent vectors, depending on the model design. 

\noindent \textbf{User and Item Tokenization.}
In conventional recommendation techniques, a user $u_i$ and an item $v_j$ are typically represented by $D$-dimensional latent vector representations, denoted as $\mathbf{p}_i$ and $\mathbf{q}_j \in \mathbb{R}^D$, respectively.
These representations can be learned with various methods, e.g., graph-based models~\cite{fan2019graph,fan2022graph}, whose effectiveness has been extensively demonstrated over the past decades.
In contrast, LLM-based recommender systems often require discrete representations for indexing users and items, in order to align with the discrete nature of language data present in their training corpora. Beyond trivial indexing methods, the most prevalent tokenization approach for LLM-based RecSys is quantization~\cite{van2017neural,rajput2023recommender,qu2024tokenrec}, which involves mapping user or item features (e.g., IDs, titles, categories, and descriptions) into a set of discrete tokens.
A vector-quantized tokenizer is typically trained via a reconstruction process, which can be formulated as follows:
\begin{align}
    \mu &= \text{Encoder} (\mathbf{x}), \\
    s &= \text{argmin}_{s} \Vert \mu - \mathbf{c}_s \Vert^2,  \\
    \hat{\mathbf{x}} &= \text{Decoder} (\mathbf{c}_s),
\end{align}
where $\mathbf{x}$ denotes the input features, $\mathbf{c}_s$ is the $s_{th}$ token in a predefined discrete token space $\mathcal{C} \in \mathbb{R}^{S \times D_c}$, and $\hat{\mathbf{x}}$ is the predicted result for the reconstruction objective.
Through this process, users $u_i$ and items $v_j$ can be tokenized using their corresponding codeword $s$ and the codeword embedding $\mathbf{c}_s$, denoted as $\mathcal{T}_i$ and $\mathcal{T}_j$ in the context of LLM-based RecSys.

\noindent \textbf{Diffusion Models.}
\label{sec:diffusion}
We provide a brief overview of diffusion models, focusing on the representative formulation known as Denoising Diffusion Probabilistic Models (DDPM)~\cite{ho2020denoising}.
In the \emph{forward process}, DDPM gradually adds Gaussian noise to the original input $\mathbf{x}^0$ according to a variance schedule $\beta_t$:
\begin{align}
    f(\mathbf{x}^{1:T}|\mathbf{x}^0) &= \prod^T_{t=1} f(\mathbf{x}^t|\mathbf{x}^{t-1}), \\
    f(\mathbf{x}^t|\mathbf{x}^{t-1}) &= \mathcal{N}(\mathbf{x}_t; \sqrt{1-\beta_t} \mathbf{x}^{t-1}, \beta_t \mathbf{I}).
    \label{eq:forward}
\end{align}
Let $\alpha_t = 1 - \beta_t$, $\overline{\alpha} = \prod^t_{\tau=1} \alpha_\tau$, $\epsilon \in \mathcal{N}(\textbf{0}, \mathbf{I})$, we can express the noisy sample at step $t$ as $\mathbf{x}^t = \sqrt{\overline{\alpha}_t} \mathbf{x}^0 + \sqrt{1-\overline{\alpha}_t} \epsilon$.
In the \emph{reverse process}, DDPM jointly models the target variable $\mathbf{x}^0$ along with a sequence of latent variables $\mathbf{x}^1$, ..., $\mathbf{x}^T$ as a Markov chain with Gaussian transitions:
\begin{align}
    g(\mathbf{x}^{0:T}) &= g(\mathbf{x}^T) \prod^T_{t=1} g(\mathbf{x}^{t-1}|\mathbf{x}^t),
\end{align}
\begin{align}
    g(\mathbf{x}^{t-1}|\mathbf{x}^t) &= \mathcal{N}(\mathbf{x}^{t-1}; \eta(\mathbf{x}^t, t), \xi(\mathbf{x}^t, t)),
\end{align}
where the initial state $\mathbf{x}^T$ is sampled from standard Gaussian noise $\mathbf{x}^T \sim \mathcal{N}(\mathbf{0}, \mathbf{I})$.
Here, $\eta(\mathbf{x}^t, t)$ denotes a time-dependent mean of the Gaussian distribution used to sample the previous timestep's data (often initialized as untrained constants), and $\xi(\mathbf{x}^t, t)$ determines the variance (or covariance) of the Gaussian distribution used in the reverse denoising process. Given the remarkable success of diffusion models in handling continuous data such as images and protein structures, an increasing number of studies have explored integrating diffusion models with LLMs~\cite{sun2024multimodal,wang2024dplm,wang2024diffusion}.
\section{Methodology}
\label{sec:method}

\subsection{Overview of \ourname{}}

\begin{figure*}
    \centering
    \includegraphics[width=0.9\linewidth]{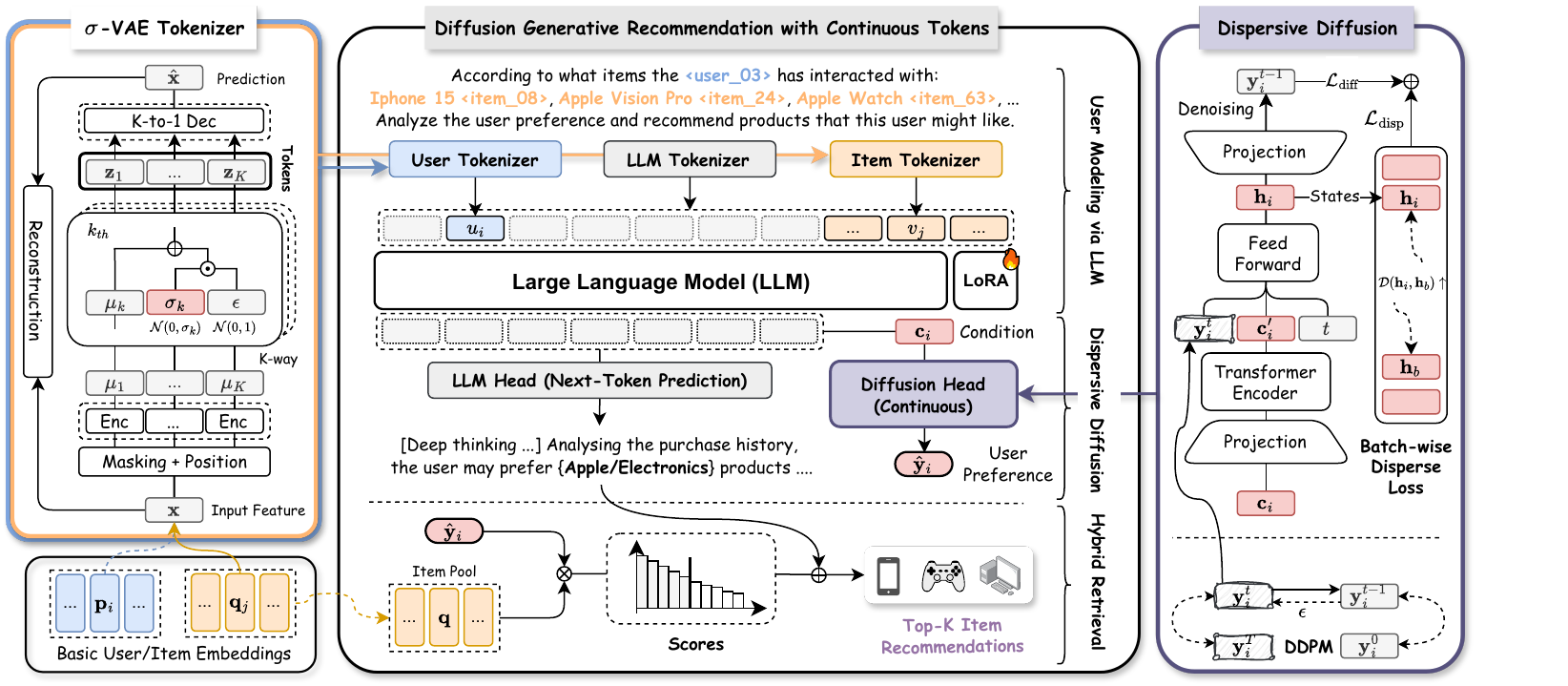}
    \captionsetup{font={small}}
        \vskip -0.1in
    \caption{Overview of the proposed \ourname{}. \ourname{} represents users\&items as latent vector representations using a not-quantized tokenizer and leverages the exceptional continuous-valued generation capability of diffusion models to operate within continuous spaces and generate implicit user preferences conditioned on the reasoning content of LLMs.}
    \label{fig:overview}
    \vskip -0.1in
\end{figure*}

In order to leverage continuous tokens into an LLM-based recommender system to enhance the model's capability for user, item, and implicit user preference representation, we propose a novel framework: \ourname{}.
As illustrated in Figure~\ref{fig:overview}, our \ourname{} consists of two core innovative modules, the \emph{$\sigma$-VAE Tokenizer} and \emph{Disperse Diffusion}.
Due to space constraints, we present only the core formulas of the key components of the proposed method in this section; supplementary information and detailed derivations are provided in Appendix~\ref{sec:app_method}.

\subsection{$\sigma$-VAE Tokenizer}
The proposed tokenizer is built upon the VAE framework~\cite{kingma2014auto,rezende2014stochastic}, which encodes input data into a latent space and subsequently decodes it back to the original space. 
Maintaining a larger variance in the latent representations improves the robustness of the VAE tokenizer to exposure bias between tokens during inference, while standard VAEs often suffer from channel variance collapse, adversely affecting autoregressive modeling~\cite{razavi2019generating}. 
To address this issue, we introduce three simple yet effective techniques to enhance our Tokenizer: (i) a masking operation based on the Bernoulli distribution to enhance robustness, (ii) a K-way encoder with parallel encoding channels to allow implicit encoding~\citep{qu2024tokenrec}, and (iii) a Gaussian kernel applied to each channel to prevent variance collapse~\cite{sun2024multimodal}.
The reconstruction pass is computed as:
\begin{align}
    \mu_k &= \text{Enc}^k(\text{Mask}(\mathbf{x}, \rho)), \\
    \mathbf{z}_k &= \mu_k + \sigma_k \odot \epsilon, \ \text{where} \ \epsilon \sim \mathcal{N}(0, 1), \ \sigma_k \sim \mathcal{N}(0, \Gamma), \\
    \label{eq:sigma}
    \hat{\mathbf{x}} &= \text{Dec}(\text{Concat}\{\mathbf{z}_k\}^K),
\end{align}
where $\sigma_k$ is a scalar coefficient sampled from a Gaussian distribution $\mathcal{N}(0, \Gamma)$, with $\Gamma$ representing the standard deviation computed from the first batch of training samples.
At its core, the $\sigma$ kernel prevents variance collapse by enforcing a fixed variance in the latent space.
In addition, the notations used in the formulas are specified as follows.
First, the input feature ($\mathbf{x}$) is basic user/item embeddings in various forms. For example, for an item, it can be the embedding derived from its textual description or a collaborative embedding learned by graph-based models.
Specifically, we set $\mathbf{x} = \mathbf{q}_j$ for item $v_j$ and $\mathbf{x} = \mathbf{p}_i$ for user $u_i$.
Second, $\text{Mask}(\cdot, \rho)$ refers to an element-wise masking strategy following a Bernoulli distribution with a masking ratio of $\rho$.
This operation generates a boolean vector of the same size as the collaborative embeddings, where each element is set to "True" with probability $\rho$ and "False" with probability $(1-\rho)$.
$\text{Enc}^k$ denotes the $k_{th}$ sub-encoder, while $\text{Dec}$ represents a K-to-1 decoder.
Both components can be implemented as three-layer Multi-Layer Perceptron (MLP) models.
Moreover, $\mu_k$ is the latent state encoded by the $k_{th}$ sub-encoder, while $\mathbf{z}_k$ is the corresponding continuous-distributed token.
In the end, each item $v_j$ or user $u_i$ (coped as $\mathbf{x}$ in the Tokenizer) can be tokenized as $\{\mathbf{z}_1, \dots, \mathbf{z}_K\}$, which are fed into the decoder to achieve the prediction value $\hat{\mathbf{x}}_j$ and trained by minimizing the reconstruction loss:
\begin{align}
    \mathcal{L}_\text{vae} = \Vert \hat{\mathbf{x}} - \mathbf{x} \Vert^2_2 + \frac{\beta}{K} \sum_{k=1}^K \Vert \mu_k \Vert^2_2.
    \label{eq:vae}
\end{align}
The first term measures the reconstruction error, and the second term is to constrain the adherence to the prior distribution~\cite{higgins2017beta}.
This constraint limits the capacity of $\mathbf{z}_k$, and encourages the model to learn the most efficient representation of the data.
The hyperparameter $\beta$ controls the trade-off between reconstruction quality and learning efficiency.

\subsection{{User Modeling via Large Language Model}}
The goal of user modeling is to capture the preference of a certain user ($u_j$) with an adequate representation.
In contrast to the majority of existing methods, which use LLMs to predict recommendation results directly, our LLM conducts user modeling by decoding user preferences in natural language and outputs the implicit user preference representation as a byproduct. 
As shown in Figure~\ref{fig:overview}, the LLM backbone summarizes the user preference (e.g., brands and categories) based on the given purchase history. 
The input $\mathcal{X}_i$ of our LLM backbone can be formed by selecting a prompt
template $\mathcal{P}$ and introducing the corresponding titles and ID tokens for user $u_i$, as well as his/her historical interacted item set $\mathcal{V}_{(u_i)}$ as follows:
\begin{align}
    \mathcal{X}_i := \mathcal{P} (\mathcal{T}_{i}, \{\mathcal{T}_{j}|v_j \in \mathcal{V}_{(u_i)}\}),
    \label{eq:prompting}
\end{align}
where $\mathcal{T}_i$ and $\mathcal{T}_j$ are the tokens used to represent user $u_i$ and item $v_j$, respectively.
In our cases, $\mathcal{T}_j$, for example, includes the title (represented by discrete and in-vocabulary tokens) of item $v_j$ and its embeddings $\{\mathbf{z}_j^1, \dots, \mathbf{z}_j^K\}$ (derived from the previous $\sigma$-VAE tokenizer module), which incorporates both discrete semantic information and continuous collaborative knowledge.
Notably, two special tokens, $\left \langle \text{z}\_\text{start} \right \rangle$ and $\left \langle \text{z}\_\text{end} \right \rangle$, are employed to denote the beginning and end of the continuous token sequence, respectively.
Please refer to Appendix~\ref{sec:app_method} for detailed prompt templates.
Following the definition of generative recommendation shown in Eq~\eqref{eq:llm4rec}, the LLM backbone models the user preference with its hybrid descriptions, with discrete and continuous tokens, by a next-token prediction format: $ \mathcal{Y}_i = \text{LLM4Rec}(\mathcal{X}_i)$. The learning objective is to minimize the following autoregressive cross-entropy loss:
\begin{align}
    \mathcal{L}_\text{llm} = - \sum^L_{l=1} \log P(\mathbf{y}_i^l | \mathbf{y}_i^{<l}, \mathcal{X}_i),
    \label{eq:llm_loss}
\end{align}
where $L$ is the maximum length of generated content, and $\mathbf{y}_i^l$ is the $l^{th}$ token of the preference reasoning texts for user $u_i$.
The LLM outputs $\mathcal{Y}_i$ may be intermediate features to generate a final recommendation (see details in Section~\ref{sec:inference}). Meanwhile, they can also be directly decoded into natural language to illustrate the reasoning process of the LLM and provide justifications for recommendation results, thereby enhancing user trust.

\subsection{{Dispersive Diffusion}}
Denoising diffusion models have recently demonstrated robust capabilities in modeling complex data distributions and learning continuous representation spaces for various modalities, including images~\citep{li2024autoregressive}, protein structures~\cite{wang2024dplm}, and videos~\citep{sun2024multimodal}. Transferring this success to LLM-based recommender systems is a promising direction to help with the delicate user preference modeling~\citep{ho2020denoising,nichol2021improved}.
To investigate the feasibility of applying diffusion models to LLM-based recommender systems, we conduct a mathematical analysis that proves the integration of continuous representations in the LLM backbone~\cite{li2024autoregressive}. 
Specifically, language models formulate the generation recommendation problem as “next token prediction” in an autoregressive manner as follows:
\begin{align}
    P(\mathbf{y}_i^1, \dots, \mathbf{y}_i^L) = \prod^{L}_{l=1}P(\mathbf{y}_i^l|\mathbf{y}_i^1,\dots,\mathbf{y}_i^{l-1}, \mathcal{X}_i).
\end{align}
A common solution is to apply the generated tokens $\{\mathbf{y}_i^1, \dots, \mathbf{y}_i^L\}$ of the LLM backbone as the representation to describe user preference. However, such a description is found to be a discrete distribution; therefore, it is less effective in modeling the complexity and subtlety of user preferences. To this end, we introduce a diffusion model conditioned on the output of the LLM backbone to further learn a continuous-distributed user preference representation space. 
Specifically, we aggregate LLM-output tokens for all previous steps to obtain the conditioning vector $\mathbf{c}_i := f_\mathbf{c}(\mathbf{y}_i^1,...,\mathbf{y}_i^{L})$.
We define the continuous-learning model as $ P(\mathbf{y}_i^{L+1}|\mathbf{c}_i)$, which can also be seen as the additional-token probability of the LLM. 
We propose parameterizing this model using diffusion models, in which the gradient can be backpropagated to $\mathbf{c}$ and further to the LLM backbone, enabling end-to-end optimization for the entire recommendation model. 

Specifically, we employ a conditional Denoising Diffusion Probabilistic Model (DDPM) with Classifier-free Guidance~\cite{ho2021classifier}. 
The forward process follows the standard DDPM formulation as presented in Eq.~\eqref{eq:forward}, while the reverse process recovers the latent representation of the target item from a standard Gaussian distribution, conditioned on $\mathbf{c}_i$.
Consider the denoising transition is intractable, the reverse process chooses to learn a model to estimate the noise $\epsilon$ given the noisy token $\mathbf{y}_i^t$ at the $t^{th}$ step.
The closed-form objective of diffusion is to minimize the following loss:
\begin{align}
    \mathcal{L}_\text{diff} &= \mathbb{E}_{(\mathbf{y}_i, \mathbf{c}_i, t)} \Vert \mathcal{F}_\theta(\mathbf{y}_i^{t}, \mathbf{c}_i, t) - \mathbf{y}_i \Vert^2_2, \\
    & = \mathbb{E}_{(\mathbf{y}_i, \mathbf{c}_i, t)} \Vert \epsilon - \epsilon_\theta (\mathbf{y}_i^t, \mathbf{c}_i, t) \Vert^2_2,
    \label{eq:diffloss}
\end{align}
where $t \in \{1, \dots, T\}$ is the sampled timestep, each corresponding to a time embedding $\mathbf{e}_t$; and $\theta$ denotes the learnable parameters of the denoising network. $\mathbf{y}_i$ is the target representation, which corresponds to the embedding of the ground-truth item, namely the target item to be recommended for the target user $u_i$. $\hat{\mathbf{y}}^t_i$ is the predicted representation during the diffusion process at $t^{th}$ step. 
As illustrated in Figure~\ref{fig:overview}, the denoising network comprises a Transformer encoder layer to encode the conditioning information, a feed-forward layer to fuse features, and two projection layers for dimensionality mapping.

The training objective typically consists of a regression term focused on reconstruction (e.g., denoising), but lacks an explicit regularization term on the representations learned for generation~\cite{yu2024repa}.
Contrastive learning provides a conceptually simple yet effective framework for learning representations from sample pairs in recommendation tasks~\cite{xie2022contrastive,cai2023lightgcl}.
It encourages attraction between item pairs interacted with by the same user (“positive pairs”) and repulsion between those that are not (“negative pairs”).
Mathematically, the typical contrastive loss in recommendation can be formulated as:
\begin{align}
    \mathcal{L}_\text{contrast} = - \log \frac{\exp(-\mathcal{D}(\hat{\mathbf{y}}_i, {\mathbf{y}}_i)/\iota)}{\sum_m \exp (-\mathcal{D}(\hat{\mathbf{y}}_i, \mathbf{y}_m)/\iota)},
    \label{eq:contrast}
\end{align}
where $(\hat{\mathbf{y}_i}, \mathbf{y}_m)$ represents sample pairs, including both the positive pair and all negative pairs.
$\mathcal{D}(\cdot)$ is the squared $\ell_2$ distance, and $\iota$ is a temperature parameter for contrastive learning, which also controls the learning speed of the diffusion process in our cases.

Despite the potential of contrastive learning as a regularizer, directly applying it would require negative sampling, which is computationally intensive in the diffusion process. 
Therefore, an emerging technique known as Dispersive Loss~\cite{wang2025diffuse} is introduced to encourage internal representations in the diffusion process to disperse within the hidden space. It is analogous to the repulsive effect of standard contrastive learning, while it requires no positive sample pairs and therefore does not interfere with the sampling process used for regression. As the original regression loss in diffusion models naturally serves as an alignment mechanism, defining positive pairs as required in standard contrastive learning would be unnecessary.
Let $h_i$ denote the internal representations of the feed-forward layer in our diffusion model for an input sample user $u_i$, we define our Dispersive Loss here as:
\begin{align}
    \mathcal{L}_\text{disp} = \log \mathbb{E}_{i,b} [\exp (- \mathcal{D}(\mathbf{h}_i, \mathbf{h}_b)/\iota)],
    \label{eq:disploss}
\end{align}
where $(\mathbf{h}_i, \mathbf{h}_b)$ represents any pair within a batch of samples. In essence, Dispersive Loss functions as a “contrastive loss without positive pairs.” Detailed derivations can be found in Appendix~\ref{sec:app_method}.

\subsection{Training and Inference}
\subsubsection{\textbf{Training}}
In our model, $\sigma$-VAE Tokenizer is responsible for mapping users and items into continuous tokens, while the hybrid LLM-Diffusion recommender backbone explores user preferences and generates personalized recommendations.
To optimize the whole model, we first train the Tokenizer using a self-supervised reconstruction objective to minimize $\mathcal{L}_\text{vae}$ as shown in Eq.~\eqref{eq:vae}.
We then train our LLM-Diffusion backbone on recommendation data and tasks, keeping the pre-trained tokenizer components fixed throughout this stage.
In this stage, three objectives are optimized jointly to build an end-to-end generative recommendation model combining the capabilities of LLMs and Diffusion.
The integrated objective is defined as:
\begin{align}
\mathcal{L} = \mathcal{L}_{\text{llm}} + \gamma_1 (\mathcal{L}_{\text{diff}} + \gamma_2 \mathcal{L}_{\text{disp}}), 
\label{eq:total_loss}
\end{align}
where $\gamma_1$ controls the trade-off between discrete output and continuous output, and $\gamma_2$ controls the impacts of the dispersive regularization.
Notably, we incorporate a small proportion of dummy conditions during the training of our denoising network to enhance the robustness and overall performance of our generative model.

\subsubsection{\textbf{Inference}}
\label{sec:inference}
For the Diffusion process~\cite{ho2021classifier} during inference, we introduce a hyperparameter $\omega$ that controls the strength of conditioning to modulate the effect of the guidance signal $\mathbf{c}_i$. We use $\omega$ to modify the noise prediction in Eq.~\eqref{eq:diffloss}.
While a higher value of $\omega$ can enhance personalized guidance, it may also reduce the generalization ability of the diffusion process, potentially deteriorating the quality of the generated oracle item.
Mahematically, the modified noise prediction process can be formulated as follows:
\begin{align}
    \tilde{\epsilon}_\theta ({\mathbf{y}}_i^t, \mathbf{c}_i, t) = (1+\omega) \epsilon_\theta ({\mathbf{y}}_i^t, \mathbf{c}_i, t) - \omega  \epsilon_\theta ({\mathbf{y}}_i^t, \Phi, t),
\end{align}
where $\Phi$ is a dummy conditioning token randomly sampled from a normal distribution.

For personalized recommendation inference, we present a \textbf{hybrid retrieval paradigm} that considers not only the textual reasoning output from LLM but also the continuous representations $\hat{\mathbf{y}}_i$ of user preference derived from a denoising network~\citep{qu2024tokenrec,fan2024survey}.
In the retrieval process, we first measure the similarity score between the $\hat{\mathbf{y}}_i$ and the predefined item representation $\hat{\mathbf{q}}_j$ for each candidate item. Meanwhile, we extract semantic information for the predicted recommendation items from the textual reasoning output of the LLM, such as category and brand. We introduce a Boolean indicator $\pi$ and combine the two parts to obtain the final similarity score for each user-item pair as follows:
\begin{align}
\label{eq:score}
    s_{ij} = \frac{\hat{\mathbf{y}}_i \mathbf{q}_j}{\Vert \hat{\mathbf{y}}_i \Vert \Vert \mathbf{q}_j \Vert} \cdot (1 + \pi).
\end{align}
For any item ($v_j$) in the candidate pool, if it is with the same semantic features as predicted by LLM, we set $\pi$ to a positive value (e.g., 0.05) and otherwise to 0. 
By ranking the items based on similarity scores calculated by Eq.~\eqref{eq:score}, \ourname{} seamlessly delivers Top-K item recommendations through a single inference process, transitioning from LLM to Diffusion.
In brief, the LLM provides a general description of user preferences, while the diffusion model generates implicit representations accordingly.

\section{Experiment}
\label{sec:experiments}
\subsection{Experimental Settings}
\subsubsection{\textbf{Datasets}}
We conducted a series of comprehensive experiments across four benchmark datasets to evaluate the effectiveness of the proposed \ourname{} and core modules in it. First, LastFM, sourced from the \emph{HetRec 2011}\footnote{\url{https://grouplens.org/datasets/hetrec-2011/}} repository, encompasses music artist listening data from users within the Last.fm online music platform.
Moreover, the ML1M dataset offers a collection of movie ratings made by \emph{MovieLens}\footnote{\url{https://grouplens.org/datasets/movielens/1m/}} users.
The last two datasets—Beauty and Video Games—are sourced from the \emph{Amazon Review Data}\footnote{\url{https://nijianmo.github.io/amazon/}} repository.
These datasets have undergone a reduction process to derive their 5-core versions, ensuring that each user and item retains a minimum of five reviews.
Table~\ref{tab:dataset} details the characteristics of these datasets, with the maximum item sequence length capped at 20 to align with a 1024-token input constraint.
In our cases, each item's title, ID, brand, and category are used, while only the ID is considered on the user side.
Moreover, our dataset division follows the Split-by-Timepoint policy~\cite{ji2023critical}, in which all interactions occurring after a predefined time point are designated as test instances, while those before the time point are used for training. 
To ensure that all items appear at least once in the training corpus, we split the recommendation data in chronological order at the 90th and 95th percentiles to create the training, validation, and testing sets.

\begin{table}[htbp]
\centering
\vskip -0.1in
\captionsetup{font={small}}
\caption{Basic statistics of benchmark datasets.}
\vskip -0.15in
\label{tab:dataset}
\begin{threeparttable}
\scalebox{0.8}{
\begin{tabular}{c|c|c|c|c|c}
\toprule
\multirow{2}{*}{\textbf{Datasets}} & \multicolumn{4}{c}{\textbf{Statistics}}           \\
                                   & \textbf{\#Users} & \textbf{\#Items} & \textbf{\#Interactions} & \textbf{Length} & \textbf{Density (\%)} \\ 
                                   \midrule
\textbf{LastFM} & 1,091  & 3,685 & 52,670 & 48.3 & 1.3101 \\
\textbf{ML1M} & 6,040 & 3,416 & 447,294 & 165.5 & 2.1679 \\
\textbf{Beauty} & 22,363 & 12,101 & 278,641 & 8.9 & 0.0734 \\
\textbf{Games} & 47,568 & 16,834 & 266,139 & 9.5 & 0.0332 \\ 
\bottomrule
\end{tabular}
}
    \end{threeparttable}
\vskip -0.1in
\end{table}

\subsubsection{\textbf{Compared Models}}
Here, we compare our approach with four representative sequential recommendation methods and five recently developed LLM-based RecSys.
The four sequential recommendation methods are \textbf{GRU4Rec}~\cite{hidasi2018recurrent}, \textbf{SASRec}~\cite{kang2018self}, \textbf{SSD4Rec}~\cite{qu2024ssd4rec}, and \textbf{DreamRec}~\cite{yang2023generate}, with DreamRec being a diffusion-based method.
The five LLM-based models are:
(i) \textbf{P5}~\cite{geng2022recommendation} reformats recommendation tasks into a text-to-text format and finetunes a small pre-trained language model (i.e., T5) to capture more profound semantics for recommendation.
(ii) \textbf{CoLLM}~\cite{zhang2023collm} incorporates collaborative representations of items into LLMs using a projection layer.
(iii) \textbf{TIGER}~\cite{rajput2023recommender} employs residual vector quantization to distill large volumes of text into a compact set of semantic identifiers, and then leverages a Transformer-based architecture trained on these semantic ID sequences to perform sequential recommendation.
(iv) \textbf{LLaRA}~\cite{liao2024llara} uses a projector to align item embeddings with textual item information and implements a curriculum learning approach for finetuning.
(v) \textbf{TokenRec}~\cite{qu2024tokenrec} proposes a novel quantized tokenizer to encode users and items along with their collaborative knowledge into discrete tokens that are compatible with LLMs.

\begin{table*}[t]
  \centering
  \captionsetup{font={small}}
  \caption{Performance comparison between representative baselines and \ourname{} across four commonly used datasets. The best and second-best results are highlighted in bold and underlined fonts, respectively. For \ourname{}, we conduct inference five times and report the mean and standard deviation. The improvements over baselines are statistically significant ($p<0.01$).}
  \vskip -0.1in
  \scalebox{0.75}{
    \begin{tabular}{cc|cccc|cccccc|c|c}
    \toprule
    \textbf{Dataset} & \textbf{Metric} & \textbf{GRU4Rec} & \textbf{SASRec} & \textbf{SSD4Rec} & \textbf{DreamRec} & \textbf{P5} & \textbf{POD} & \textbf{CoLLM} & \textbf{TIGER*} & \textbf{TokenRec} & \textbf{LLaRA} & \textbf{\ourname{}} & \textbf{Improve.*} \\
    \midrule
    \multirow{4}[2]{*}{\textbf{Beauty}} & HR@10 & 0.0394  & 0.0407  & 0.0412  & 0.0423  & 0.0411  & 0.0414  & 0.0429  & 0.0439  & 0.0438  & \underline{0.0442}  & \textbf{0.0473}$_{\pm0.0017}$  & 7.7449\% \\
          & HR@20 & 0.0595  & 0.0611  & 0.0628  & 0.0642  & 0.0594  & 0.0606  & 0.0609  & 0.0620  & 0.0619  & \underline{0.0631}  & \textbf{0.0669}$_{\pm0.0024}$  & 7.9032\% \\
          & NDCG@10 & 0.0204  & 0.0211  & 0.0218  & 0.0222  & 0.0236  & 0.0227  & 0.0237  & 0.0243  & 0.0226  & \underline{0.0237}  & \textbf{0.0251}$_{\pm0.0011}$   & 3.2922\% \\
          & NDCG@20 & 0.0253  & 0.0258  & 0.0256  & 0.0265  & 0.0258  & 0.0254  & 0.0266  & 0.0269  & 0.0273  & \underline{0.0283}  & \textbf{0.0295}$_{\pm0.0012}$   & 9.6654\% \\
    \midrule
    \multirow{4}[2]{*}{\textbf{Games}} & HR@10 & 0.0732  & 0.0783  & 0.0875  & 0.0915  & 0.0839  & 0.0858  & 0.0928  & 0.0958  & \underline{0.1018}  & 0.0997  & \textbf{0.1041}$_{\pm0.0036}$  & 8.6639\% \\
          & HR@20 & 0.1139  & 0.1231  & 0.1345  & 0.1427  & 0.1269  & 0.1286  & 0.1398  & 0.1438  & \underline{0.1506}  & 0.1471  & \textbf{0.1606}$_{\pm0.0074}$   & 11.6829\% \\
          & NDCG@10 & 0.0365  & 0.0447  & 0.0468  & 0.0489  & 0.0430  & 0.0426  & 0.0491  & 0.0546  & \underline{0.0570}  & 0.0568  & \textbf{0.0588}$_{\pm0.0017}$  & 7.6923\% \\
          & NDCG@20 & 0.0467  & 0.0536  & 0.0545  & 0.0626  & 0.0554  & 0.0542  & 0.0618  & 0.0679  & \underline{0.0705}  & 0.0690  & \textbf{0.0704}$_{\pm0.0013}$  & 3.6819\% \\
    \midrule
    \multirow{4}[2]{*}{\textbf{LastFM}} & HR@10 & 0.0340  & 0.0353  & 0.0358  & 0.0359  & 0.0386  & 0.0402  & 0.0468  & 0.0467  & 0.0489  & \underline{0.0525}  & \textbf{0.0539}$_{\pm0.0034}$   & 15.4176\% \\
          & HR@20 & 0.0467  & 0.0470  & 0.0469  & 0.0493  & 0.0555  & 0.0672  & 0.0732  & 0.0749  & 0.0755  & \underline{0.0827}  & \textbf{0.0890}$_{\pm0.0067}$   & 18.8251\% \\
          & NDCG@10 & 0.0191  & 0.0202  & 0.0195  & 0.0198  & 0.0190  & 0.0219  & 0.0228  & 0.0226  & 0.0238  & \underline{0.0244}  & \textbf{0.0250}$_{\pm0.0010}$  & 10.6195\% \\
          & NDCG@20 & 0.0239  & 0.0237  & 0.0235  & 0.0244  & 0.0225  & 0.0268  & 0.0305  & 0.0306  & 0.0311  & \underline{0.0325}  & \textbf{0.0338}$_{\pm0.0015}$   & 10.4575\% \\
    \midrule
    \multirow{4}[2]{*}{\textbf{ML1M}} & HR@10 & 0.0877  & 0.0964  & 0.0951  & 0.0944  & 0.0867  & 0.0886  & 0.0923  & 0.0954  & \underline{0.1076}  & {0.1012}  & \textbf{0.1099}$_{\pm0.0066}$  & 15.1992\% \\
          & HR@20 & 0.1250  & 0.1382  & 0.1345  & 0.1371  & 0.1248  & 0.1277  & 0.1499  & 0.1548  & 0.1506  & \textbf{0.1672}  & \underline{0.1666}$_{\pm0.0084}$  & 7.6227\% \\
          & NDCG@10 & 0.0379  & 0.0399  & 0.0396  & 0.0403  & 0.0381  & 0.0373  & 0.0456  & 0.0472  & \underline{0.0542} & 0.0532  & \textbf{0.0561}$_{\pm0.0039}$  & 18.8559\% \\
          & NDCG@20 & 0.0487  & 0.0510  & 0.0511  & 0.0540  & 0.0486  & 0.0487  & 0.0620  & 0.0644  & \underline{0.0716}  & 0.0698  & \textbf{0.0717}$_{\pm0.0035}$  & 11.3354\% \\
    \bottomrule
    \end{tabular}%
    }
  \label{tab:comparison}%
  \vskip -0.1in
\end{table*}%
\subsubsection{\textbf{Evaluation Metrics}}
Two commonly used metrics are used: Top-K Hit Ratio (HR@K) and Top-K Normalized Discounted Cumulative Gain (NDCG@K)~\citep{he2020lightgcn}, where higher values indicate superior performance.
The evaluation entails presenting the average metrics for all users in the test set.
Furthermore, the values of K are specified as 10 and 20, with 10 serving as the default setting for ablation experiments and parameter analyses.

\subsubsection{\textbf{Configurations}}
We implement all models using Python 3.11 and PyTorch 2.5.1 on an NVIDIA A800 (80 GB) GPU.
We use AdamW with learning rates and weight decays of 0.00001/0.0001 for the backbone and 0.0001/0.001 for the tokenizer.
The maximum length of the item sequence is configured to 20.
The masking ratio $\rho$ is searched from 0 to 1 in 0.2 increments, while the number of sub-encoders is selected from \{1, 2, 3, 4, 5\}.
Taking into account both performance and computational resources, we opt to utilize the \emph{Llama-3.2-1B-Instruct}\footnote{\url{https://huggingface.co/meta-llama/Llama-3.2-1B-Instruct}} as our LLM backbone, which is renowned as one of the most popular pre-trained language models globally. The LLM backbone operates with a default temperature of 0.
We create ten prompt templates and three response templates for the input and output of the LLM backbone, respectively.
A random selection from these templates is made in each batch to enhance the model's capacity to generalize across various template structures.
The two loss weights $\gamma_1$ and $\gamma_2$ are sequential searched in the range of \{0, 0.5, 1.0, 1.5, 2.0\}.
The basic embeddings for users and items are derived from both textual information encoded by SentenceBERT~\cite{reimers2019sentence} and collaborative knowledge learned through LightGCN~\citep{he2020lightgcn}.
Our noise schedule consists of 1000 steps during training; during inference, it is resampled with fewer steps (100).
We fix the unconditional training probability $\zeta$ as 0.1 suggested by \cite{ho2021classifier}, and the diffusion temperature $\iota$ to 0.5 following~\cite{wang2025diffuse}.
We search for the personalized guidance strength $\omega$ in the range of \{1, 2, 4, 6, 8, 10\}.
More implementation details are presented in Appendix~\ref{sec:app_result}.

\subsection{Recommendation Performance}

Table~\ref{tab:comparison} presents the overall performance comparison between \ourname{} and the baselines over four investigated datasets.
From the table, we have the following observations:
\begin{itemize}[leftmargin=*]
    \item The proposed model achieves superior performance, outperforming all representative baselines in the majority of cases.
    On average, \ourname{} significantly exceeds the typical baseline with discrete tokens (TIGER) by 11.76\% on HR@10 and 10.11\% on NCDG@10 across the four datasets.
    When compared to the strongest baseline (TokenRec and LLaRA), the enhancements amount to 5.67\% and 4.33\% correspondingly.
    Such improvement demonstrates the effectiveness of our proposed method and the great potential of exploring diffusion continuous tokens in LLM-based recommender systems.
    \item Leveraging its exceptional reasoning capability developed from million/billion-level parameters and extensive training data, the LLM-based methods, with personalized fine-tuning on these datasets, empirically surpass the collaborative filtering methods and sequential recommendation models.
    \item Among the LLM-based models, those incorporating continuous tokens for representing users/items (i.e., CoLLM and LlaRA) outperform the discrete method (i.e., P5 and TIGER), indicating the advantage of using continuous tokens.
    Nevertheless, the inherent disparity between discrete and continuous token spaces impedes the performance of these models, as evidenced by the improved outcomes achieved by TokenRec, which introduces an advanced tokenizer to index users and items with several discrete tokens.
    To operate with continuous tokens, our method adopts the proposed continuous tokenizer and dispersive diffusion, enabling LLM-based RecSys to benefit from non-quantized user/item representations and attain higher-quality generation.
    \item Additionally, among deep learning methods, DreamRec outperforms traditional sequential recommendation approaches such as GRU4Rec, SASRec, and SSD4Rec. These results highlight the effectiveness of diffusion models in generating precise user preferences through step-by-step denoising diffusion.
\end{itemize}

\subsection{In-depth Analysis}

\subsubsection{\textbf{Model Optimization Robustness Analysis}}
\begin{figure}[t]
    \centering
    \includegraphics[width=0.8\linewidth]{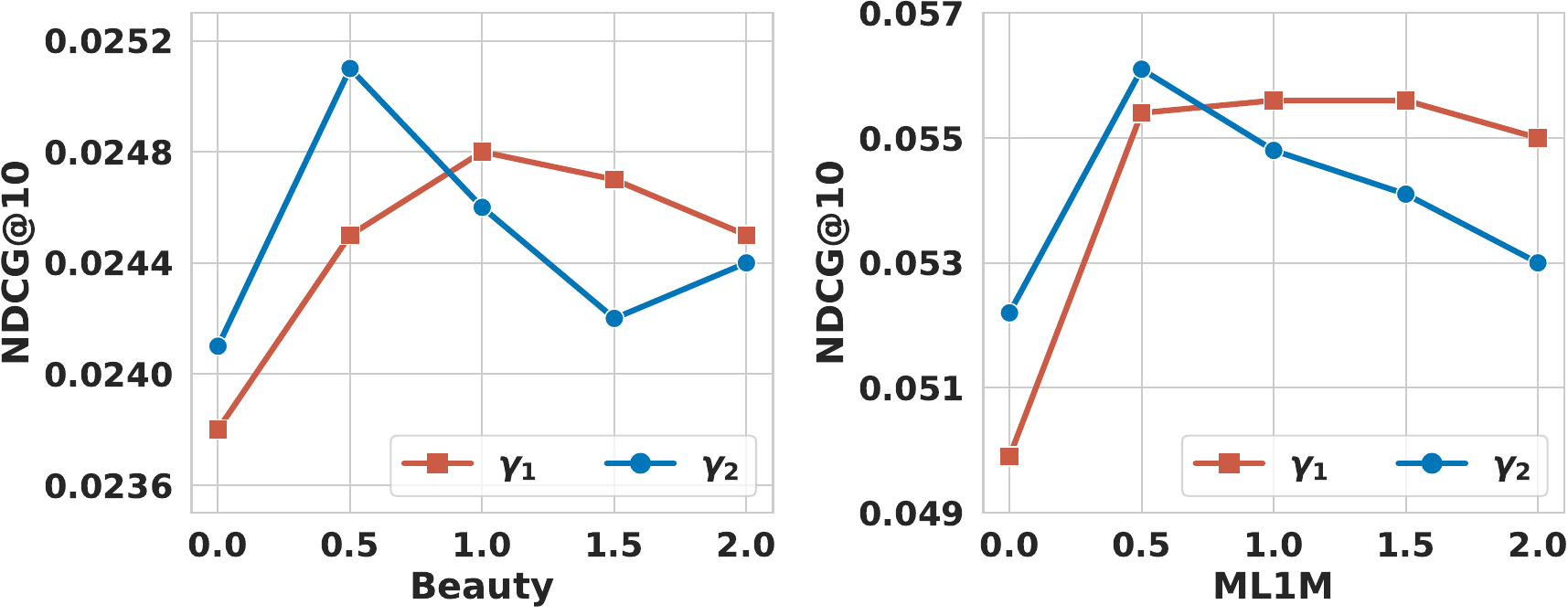}
    \vskip -0.1in
      \captionsetup{font={small}}
    \caption{Optimization robustness on the Beauty and ML1M datasets.}
    \label{fig:sensitivity}
    \vskip -0.1in
\end{figure}

We assess the sensitivity of the optimization process of \ourname{}’s with regard to two hyperparameters, as defined in the end-to-end learning loss in Eq.~\eqref{eq:total_loss}.
$\gamma_1$ controls the trade-off between LLM and Diffusion parts, and $\gamma_2$ determines the impact of dispersive loss specifically in the Diffusion.
Using a sequential grid search over \{0, 0.5, 1, 1.5, 2\}, we first vary $\gamma_1$ with $\gamma_2$ fixed at 0, then assess the effect of $\gamma_2$ using the optimal $\gamma_1$ from the first step.
As illustrated in Figure~\ref{fig:sensitivity}, the results show that the model performance stays relatively stable with the shift of the two hyperparameters. 
This demonstrates the robustness of \ourname{}’s learning process.
Moreover, optimal performance for $\gamma_1$ is achieved when its value is set to 1, while the optimal value for $\gamma_2$ is 0.5.

\subsubsection{\textbf{Effect of Key Hyperparameters}}
\begin{figure}[t]
    \centering
    \includegraphics[width=0.9\linewidth]{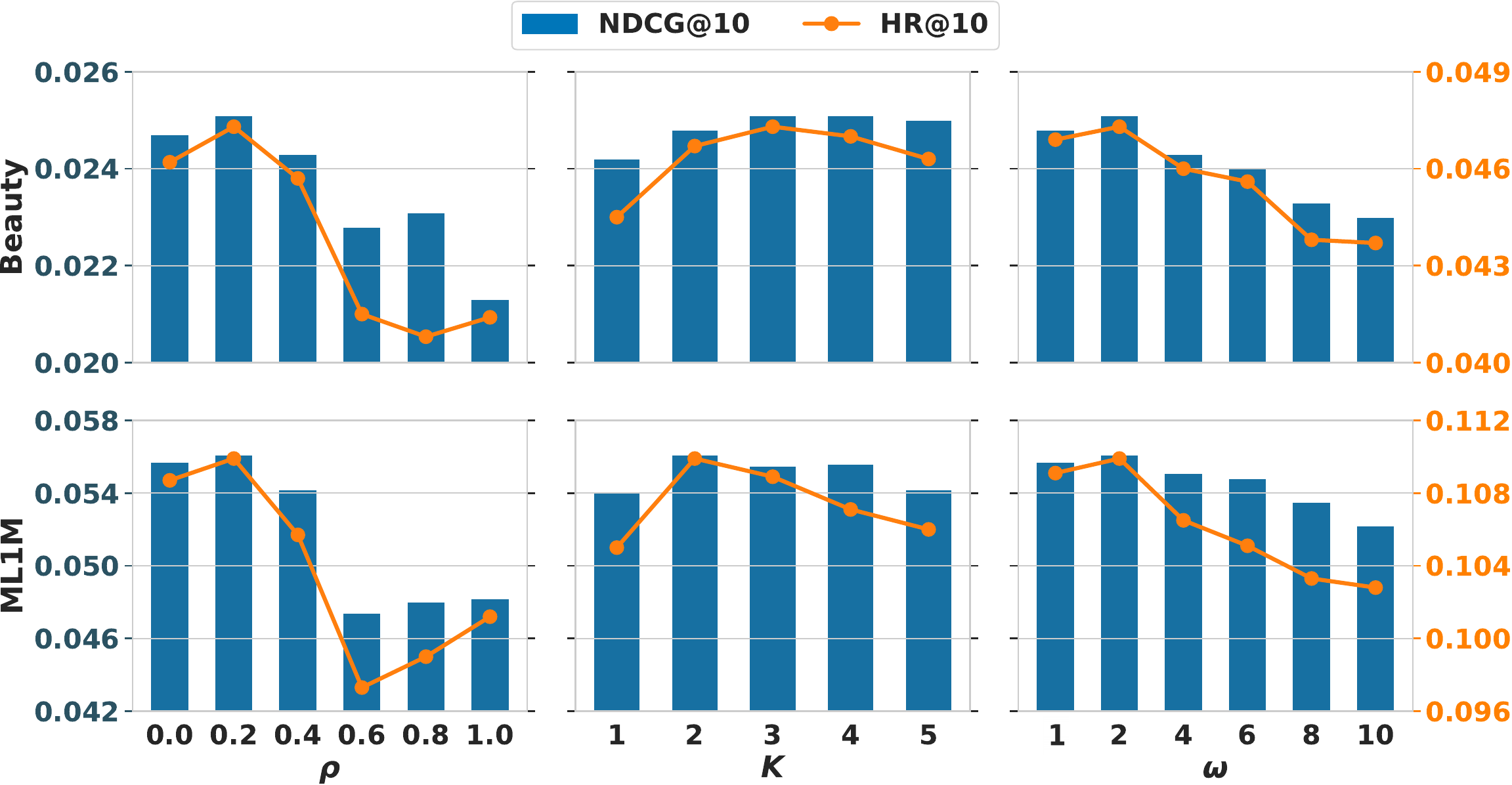}
    \vskip -0.1in
      \captionsetup{font={small}}
    \caption{Hyper-parameter tuning on the Beauty and ML1M datasets.}
    \label{fig:param}
    \vskip -0.1in
\end{figure}

This section examines three critical hyperparameters in our approach: the classifier-free guidance $\omega$, the masking ratio $\rho$, and the token number $K$ for each user/item.
Figure~\ref{fig:param} illustrates the NDCG@10 and HR@10 change observed in \ourname{}.
First, the hyperparameter $\rho$ controls the masking ratio of our encoder for robustly tokenizing users and items with continuous tokens. The results show that minor incorporation of masking during training yields performance enhancements, with a ratio of 0.2 emerging as the optimal value in our experiments. Furthermore, our experimental findings demonstrate that the recommendation performance deteriorates for masking ratios $\rho \geq 0.4$, underscoring the importance of avoiding excessive masking practices.
Second, the figure shows that indexing users and items with three or four tokens yields the best results for datasets such as Beauty, while two tokens are sufficient for ML1M due to its relatively smaller number of users and items. This observation is generally consistent with findings reported in previous studies, e.g., TIGER~\cite{rajput2023recommender}, TokenRec~\cite{qu2024tokenrec}, and OneRec~\cite{deng2025onerec}.
Finally, Classifier-free guidance~\citep{ho2021classifier} is widely used in diffusion models to balance fidelity and realism by scaling the difference between conditional and unconditional predictions.
Following previous studies~\cite{li2024autoregressive,sun2024multimodal,luo2025deem}, we search the hyperparameter $\omega$ in the range of \{0, 2, 4, 6, 8, 10\}.
The results show that employing a small CFG ($\omega=2$) improves overall performance across both datasets, while larger values may lead to degradation in output quality.

\subsubsection{\textbf{Reconstruction Evaluation}}
\label{sec:reconstruction}
The reconstruction performance of a model largely reflects the quality of its learned representation, which in turn affects its performance on downstream tasks~\cite{van2017neural,chen2025softvq}, such as the item/user tokenizer in our recommendation task.
To investigate the effectiveness of continuous tokens in recommendation scenarios, we evaluate various methods for the item embedding reconstruction task on the Beauty dataset and report all the performance in Figure~\ref{fig:reconstruction}.
In our experimental setting, each model takes the item embeddings learned by LightGCN~\cite{he2020lightgcn} as input and encodes them into tokens, which are then decoded to reconstruct the input (for diffusion models, this reconstruction process includes adding and removing noise).
We can see that, although advanced quantization-based methods (RQ- and MQ-VAE) demonstrate improvements compared to VAE and VQ-VAE, methods with continuous tokens and decoding using the diffusion model achieve superior reconstruction performance and loss convergence.
This can be attributed to the ability of continuous tokens and diffusion models to model subtle differences.

\begin{table}[t]
  \centering
  \captionsetup{font={small}}
  \caption{Results of ablation studies. NDCG@10 is abbreviated as NG@10 throughout the table.}
  \vskip-0.1in
  \scalebox{0.7}{
    \begin{tabular}{cc|cccc}
    \toprule
    \multirow{2}[1]{*}{\textbf{Module}} & \multirow{2}[1]{*}{\textbf{Conponent}} & \multicolumn{2}{c}{\textbf{Beauty}} & \multicolumn{2}{c}{\textbf{ML1M}} \\
          &       & \textbf{HR@10} & \textbf{NG@10} & \textbf{HR@10} & \textbf{NG@10} \\
    \midrule
    ContRec & Full  & 0.0473  & 0.0251  & 0.1099  & 0.0561  \\
    \midrule
    \multirow{5}[4]{*}{Tokenization} & w/o Masking & 0.0462  & 0.0247  & 0.1087  & 0.0557  \\
          & w/o User & 0.0461 & 0.0246 & 0.1066 & 0.0555 \\
          & w/o $\sigma$ & 0.0457  & 0.0242  & 0.1051  & 0.0536  \\
          & w/o Tokenizer & 0.0438  & 0.0237  & 0.1018  & 0.0511  \\
          & w/o Text & 0.0456 & 0.0242 & 0.1070 & 0.0550 \\
          & w/o Collaborative & 0.0448 & 0.0236 & 0.1030 & 0.0534 \\
\cmidrule{2-6}          & w/ RQ-VAE & 0.0438  & 0.0236  & 0.0996  & 0.0508  \\
          & w/ VQ-VAE & 0.0426  & 0.0236  & 0.0974  & 0.0488  \\
    \midrule
    \multirow{3}[2]{*}{Generation} & w/o $\pi$ & 0.0463  & 0.0244  & 0.1063  & 0.0544  \\
          & w/o $\mathcal{L}_\text{disp}$ & 0.0448  & 0.0241  & 0.1042  & 0.0522  \\
          & w/o Diffusion & 0.0431  & 0.0238  & 0.1007  & 0.0499  \\
    \bottomrule
    \end{tabular}%
    }
  \label{tab:ablation}%
  \vskip -0.1in
\end{table}%

\subsection{Ablation Study}
We conducted ablation experiments on the Beauty and ML1M datasets, where each key component was eliminated separately as follows:
(i) w/o Masking: The masking operation in the encoder is deactivated.
(ii) w/o User: The user tokens are removed, leaving only item tokens in our prompt.
(iii) w/o $\sigma$: The regularization coefficient in Eq.~\eqref{eq:sigma} is removed.
(iv) w/o Tokenizer: The proposed tokenizer is replaced with a projection layer.
(v) w/o Text: The basic embeddings exclude textual item-side information.
(vi) w/o Collaborative: The basic embeddings are constructed without collaborative embeddings.
(vii) w/ RQ-VAE: The continuous $\sigma$-VAE is substituted with a discrete RQ-VAE.
(viii) w/ VQ-VAE: The $\sigma$-VAE is replaced with a vanilla VQ-VAE.
(viv) w/o $\pi$: The LLM reasoning result is excluded from the retrieval process.
(vv) w/o $\mathcal{L}_\text{disp}$: The dispersive loss is removed from the overall objective function in Eq.~\eqref{eq:total_loss}.
(vvi) w/o Diffusion: The diffusion head is removed and replaced with a projection layer.
From the ablation results presented in Table~\ref{tab:ablation}, we can draw several key observations.
First, all major components of our approach contribute to overall performance, as removing any single component leads to a decrease in effectiveness.
Second, the continuous $\sigma$-VAE consistently outperforms its discrete counterparts, RQ-VAE and VQ-VAE.
Finally, eliminating either the dispersive loss or the diffusion module results in a substantial drop in performance, highlighting the critical importance of the proposed diffusion-based continuous token generation in generative recommendation systems.
Further results, including inference costs, case analyses, and generalizability, are provided in Appendix~\ref{sec:app_result}.

\section{Conclusion}
\label{sec:conclusion}
To overcome the limitations of lossy tokenization and suboptimal learning in quantization methods, we introduced \ourname{}, a novel framework that integrates continuous tokenization into LLM-based recommender systems.
By leveraging a $\sigma$-VAE Tokenizer and a Dispersive Diffusion module, ContRec effectively encodes users and items as continuous tokens and captures implicit user preferences through conditional diffusion processes.
Extensive experiments on four benchmark datasets demonstrate that ContRec consistently outperforms both traditional and state-of-the-art LLM-based recommender systems.
These results highlight the significant potential of continuous tokenization and generative modeling in advancing the next generation of recommender systems.
\begin{acks}
This work was supported by the National Natural Science Foundation of China (project No.62406336). 
The research described in this paper has been partially supported by the General Research Funds from the Hong Kong Research Grants Council (project no. PolyU 15207322, 15200023, 15206024, and 15224524), 
internal research funds from Hong Kong Polytechnic University (project no. P0042693, P0048625, and P0051361), and Sheertek International (HK) Limited.   
This work was supported by computational resources provided by The Centre for Large AI Models (CLAIM) of The Hong Kong Polytechnic University.
\end{acks}

\vfill\eject
\bibliographystyle{ACM-Reference-Format}
\balance
\bibliography{sample-base}

\appendix
\section{Related Work}
\label{sec:literature}

\subsection{LLM-based Recommender Systems}
The rapid advancement of Large Language Models (LLMs), such as ChatGPT and GPT-4, has significantly transformed natural language processing~\citep{ning2025survey,fan2025computational,qu2024survey,kaplan2020scaling}.
As a result, LLMs have increasingly been integrated into recommender systems, enhancing their capabilities~\citep{wang2025knowledge,zhao2024recommender,huang2025towards}.
To effectively leverage LLMs in recommendation tasks, two primary challenges must be addressed: the tokenization of users and items, and the generation of recommendations. Most research has focused on adapting recommender systems to the language domain using discretization techniques, which align with the discrete nature of natural language.
A notable example is P5~\citep{geng2022recommendation}, which unifies various recommendation tasks into natural language processing through multi-task prompt-based pre-training, achieving impressive zero-shot generalization with personalized prompts.
Following this, simple tokenization methods like title indexing, independent indexing, and sequential indexing have become common~\citep{hua2023index}, as seen in studies like TallRec~\citep{bao2023tallrec} and LLMRec~\citep{wei2024llmrec}.
However, these methods struggle to capture complex collaborative knowledge and are limited by LLMs' input length constraints.
To overcome these limitations, researchers have developed robust tokenizers that convert users, items, and their interactions into discrete tokens.
Examples include TIGER~\citep{rajput2023recommender} and TokenRec~\citep{qu2024tokenrec}, which use vector-quantized tokenizers to encode textual information and collaborative representations from graph neural network-based methods~\citep{he2020lightgcn}. Despite the performance improvements offered by advanced discrete-valued tokenizers, they are hindered by information loss during compression or quantization.
Consequently, there is growing interest in continuous-valued tokens, which have shown promise in representation learning~\citep{yue2023llamarec,zhang2023collm,liao2024llara}.
However, this approach is still in its early stages of development.
Such a direct approach faces challenges in bridging the substantial gap between the continuous collaborative knowledge that GNNs capture from user-item interactions and the semantic information that LLMs learn from discrete natural language.

\subsection{Diffusion Models with Continuous Tokens}
Denoising diffusion probabilistic models, also known as DDPM or diffusion models, function based on the concept of learning to reverse a gradual noise-addition process~\citep{ho2020denoising,nichol2021improved}.
In contrast to language models that commonly operate with discrete tokens, diffusion models work with continuous vectors, rendering them well-suited for tasks that involve continuous data, such as images.
For example, palette~\citep{saharia2022palette} demonstrates strong performance across a range of image-to-image generation tasks by employing a general diffusion model, eliminating the need for task-specific fine-tuning.
Given the success of diffusion models in continuous data generation, an increasing number of studies have proposed combining large language models with diffusion models in cross-modal modeling, with both continuous and discrete data as inputs and targets.
Examples include VAE-MAR~\citep{li2024autoregressive} and Next-Token Diffusion~\citep{sun2024multimodal}, which employ a variational autoencoder (VAE) to represent continuous image data as latent vectors and introduce denoising diffusion for autoregressive generation of these vectors.
In these studies, language models continue their imposition on discrete language generation and comprehension, while diffusion models are incorporated as generation heads to operate within continuous domains.
Despite that, the integration of diffusion models with auto-regressive models is currently in its nascent phase.
\section{Further Explanation of \ourname{}}
\label{sec:app_method}

\subsection{Algorithm of the Proposed Framework}
The pseudo-code of the proposed framework during training and inference is described in Algorithm~\ref{alg:inference}.
The ultimate goal is to develop a comprehensive understanding of user preferences for generating effective and accurate recommendations.
Top-K relevant items can be retrieved from the ranking list as the recommended output.

\begin{algorithm}[!h]
    \caption{\ourname{}: Pseudo-code.}
    \label{alg:inference}
    \renewcommand{\algorithmicensure}{\textbf{Output:}}
    \begin{algorithmic}[1]
        \renewcommand{\algorithmicrequire}{\textbf{Init:}}
        \REQUIRE Embeddings bases for users $\{\mathbf{p}_i\}^n$ and items $\{\mathbf{q}_j\}^m$
        
        \renewcommand{\algorithmicrequire}{\textbf{Training:}}
        \REQUIRE

        \renewcommand{\algorithmicrequire}{\textbf{Phase1:}}
        \REQUIRE $\sigma$-VAE Tokenizer
        \FOR{each $\mathbf{p}_i$ or $\mathbf{q}_j$}
        \STATE Encoding: $\mu_k = \text{Enc}^k(\text{Mask}(\mathbf{x}, \rho))$;
        \STATE Not-quantized tokenization: $\mathbf{z}_k = \mu_k + \sigma_k \odot \epsilon$;
        \STATE Decoding: $\hat{\mathbf{x}}=\text{Dec}(\text{Concat}\{\mathbf{z}_k\}^K)$;
        \STATE Calculating the reconstruction loss based on Eq.~\eqref{eq:vae}.
        \ENDFOR
        \RETURN $K$ continuous tokens for each user and item.
        \renewcommand{\algorithmicrequire}{\textbf{Phase2:}}
        \REQUIRE LLM and Diffusion
        \FOR{each $u_i$ with item interactions $\mathcal{V}_{u_i}$}
        \STATE Prompting to $\mathcal{X}_i$ as expressed in Eq.~\eqref{eq:prompting};
        \STATE Auto-regressive generation via LLM: $\mathcal{Y}_i = \text{LLM}(\mathcal{X}_i)$;
        \STATE Continuous generation via Diffusion: $\mathbf{y}_i=\text{Diffusion}(\mathbf{c}_i, \epsilon)$;
        \STATE Gradient backpropagation through Eq.~\eqref{eq:total_loss}.
        \ENDFOR

        \STATE -----------------------------------------------------------------------------------------
        \renewcommand{\algorithmicrequire}{\textbf{Inference:}}
        \REQUIRE

        \FOR{each $u_i \in \mathcal{U}$}
        \STATE $\mathcal{T}_j: (|\mathcal{V}_{u_i}|, K, D) =\{ \{{\mathbf{z}}^{v_j}_k\}^K| v_j\in\mathcal{V}_{u_i} \} \xleftarrow{} \text{Tokenizer}(\mathbf{q}_j)$;
        \STATE $\mathcal{T}_i: (K, D) =\{{\mathbf{z}}^u_k\}^K \xleftarrow{} \text{Tokenizer}(\mathbf{p}_i)$;
        \STATE Prompting $\mathcal{X}_i: (L_{in}, D) \xleftarrow{} \mathcal{P}({\mathcal{T}}_i, \{{\mathcal{T}}_j | v_j \in \mathcal{V}_{u_i} \})$;
        \STATE Reasoning $\mathcal{Y}_i: (L, D) = \text{LLM}(\mathcal{X}_i)$;
        \STATE Conditioning $\mathbf{c}_i: (D) = f(\mathbf{y}_i^1,\dots,\mathbf{y}_i^{L-1})$;
        \STATE Denoising $\mathbf{y}_i: (d) = \text{Diffusion}(\mathbf{c}_i, \epsilon)$;
        \STATE Hybrid scoring $s_{ij} \xleftarrow{}$ Eq.\eqref{eq:score};
        \ENDFOR
        \RETURN Scores for all users towards all items $\{s_{ij}\} \in \mathbb{R}^{|\mathcal{U}| \times |\mathcal{V}|}$.
    \end{algorithmic}
\end{algorithm}

\subsection{Derivation of Dispersive Loss}
In our case, we derive the dispersive loss~\cite{wang2025diffuse} from the typical contrastive learning loss, namely InfoNCE loss.
Formally, the original InfoNCE loss~\cite{oord2018representation} can be interpreted as a categorical cross-entropy objective that encourages high similarity between positive pairs and low similarity between negative pairs, as shown in Eq.~\eqref{eq:contrast}.
We can rewrite this equation as:
\begin{align}
\mathcal{L}_\text{contrast}=\mathcal{D}(\hat{\mathbf{y}}_i, \mathbf{y}_i)/\tau + \log \sum_m \exp (-\mathcal{D}(\hat{\mathbf{y}}_i, \mathbf{y}_m)/\tau),
\label{eq:contrast2}
\end{align}
where the numerator involves only the positive pair, whereas the denominator includes all pairs in the batch.
Here, the first term is similar to a regression objective, which minimizes the distance between the prediction $\hat{\mathbf{y}}_i$ and its target $\mathbf{y}_i$.
On the other hand, the second term encourages any pair to be as distant as possible.

To construct Dispersive Loss, we conduct the training on the intermediate representations $\mathbf{h}_i$ of the generative model for user $u_i$ as a replacement of $\mathbf{y}_i$, followed by~\cite{wang2025diffuse}.
Building upon this, we keep only the second term in Eq~\eqref{eq:contrast2}:
\begin{align}
    \mathcal{L}_\text{disp}=\log \sum_m \exp(-\mathcal{D}(\mathbf{h}_i, \mathbf{h}_m)/\tau).
\end{align}
Moreover, to have a form defined on a batch of samples, the formula can be redefined as:
\begin{align}
    \mathcal{L}_\text{disp}=\log \mathbb{E}_{i,b}[\exp (-\mathcal{D}(\mathbf{h}_i, \mathbf{h}_b)/\tau)].
\end{align}
This loss function yields the same value for all samples within a batch and is computed only once per batch.
By removing the alignment term on the intermediate representations, it further emphasizes the regularization aspect.
Notably, the loss does not need to explicitly exclude the pair $\mathcal{D}(\mathbf{h}_i, \mathbf{h}_b)$ when $i=b$, as it maintains a unified view of all samples in each batch.

\subsection{Prompt Templates}
Inspired by the success of in-context learning in large language models, we employ two widely used prompting techniques, namely role play and few-shot learning.
Let "$\left \langle \cdot \right \rangle$" represents a continuous encoded token for users or items. A detailed example of our prompt template is outlined in the colorbox.
Please refer to our code for the detailed prompts.

\begin{tcolorbox} 
$\Longrightarrow$ \textbf{In-context Learning}:  \\
You are the expert in \{dataset\} products, entrusted with the responsibility of recommending the perfect products to our users. \\

Here is an example format for recommendations: \\

\#\#\# Input: Given the following purchase history: [item side information]. I wonder what the user will like. Can you help me decide? \\
\#\#\# Response: The interaction history shows that the user might like [item\_brand/item\_category] products.' \\

Now, please provide your recommendations based on the following content:
\tcblower
$\Longrightarrow$ \textbf{Sample Content}: \\
\#\#\# Input: Considering the user

\colorbox{cyan!30}{$ \left \langle \text{z}\_\text{start} \right \rangle \left \langle u_i^1\text{-emb} \right \rangle \left \langle u_i^2\text{-emb} \right \rangle\left \langle u_i^3\text{-emb} \right \rangle \left \langle \text{z}\_\text{end} \right \rangle$}

has interacted with:

\colorbox{red!30}{$itemtitle\ \left \langle \text{z}\_\text{start} \right \rangle \left \langle v_1^1\text{-emb} \right \rangle \left \langle v_1^2\text{-emb} \right \rangle\left \langle v_1^3\text{-emb} \right \rangle \left \langle \text{z}\_\text{end} \right \rangle $}, 
\colorbox{red!30}{$itemtitle\ \left \langle \text{z}\_\text{start} \right \rangle \left \langle v_2^1\text{-emb} \right \rangle \left \langle v_2^2\text{-emb} \right \rangle\left \langle v_2^3\text{-emb} \right \rangle \left \langle \text{z}\_\text{end} \right \rangle$},
\colorbox{red!30}{$itemtitle\ \left \langle \text{z}\_\text{start} \right \rangle \left \langle v_3^1\text{-emb} \right \rangle \left \langle v_3^2\text{-emb} \right \rangle\left \langle v_3^3\text{-emb} \right \rangle \left \langle \text{z}\_\text{end} \right \rangle$},

what are the expected preferences of the user? \\
\#\#\# Response:
\end{tcolorbox}

\section{Further Analysis of Experimental Results}
\label{sec:app_result}

\subsection{Analysis of Success and Failure Cases.}
Table~\ref{tab:cases} shows two success cases (users 2715 and 36183) and a failure case (user 7849) of \ourname{}.  Notably, the failure case for user 7849 demonstrates a scenario in which \ourname{} makes an incorrect prediction due to the user's shift in type, platform, and category preferences. This highlights a potential limitation of our model: the LLM-based reasoning tends to favor items with categories similar to those in the user's historical interactions, making it challenging to accommodate sudden shifts in user preferences.

\begin{table*}[t]
  \centering
      \captionsetup{font={small}}
  \caption{Examples of two success cases (users 2715 and 36183) and a failure case (user 7849). The target item is underlined.} \vskip -0.1in
  \scalebox{0.9}{
    \begin{tabular}{cccccc}
    \toprule
    \textbf{User ID} & \textbf{Item Sequence $\downarrow$} & \textbf{Title} & \textbf{Type} & \textbf{Platform} & \textbf{Category} \\
    \midrule
    \multirow{5}[1]{*}{2715 (S)} & 523   & Namco Museum 64 & Retro Gaming \& Microconsoles & Nintendo 64 & Games \\
          & 552   & Jungle Gree & Retro Gaming \& Microconsoles & Nintendo 64 & Games \\
          & 6386  & Video Cable Cord & Retro Gaming \& Microconsoles & Trenro & Accessories \\
          & 12289 & OSTENT Wired Controller Gamepad Joystick Joypad & Retro Gaming \& Microconsoles & Nintendo 64 & Accessories \\
          & \underline{9142}  & Booster Pack (Jumper Pak) & Retro Gaming \& Microconsoles & Nintendo 64 & Accessories \\
                    \midrule
    \multirow{5}[0]{*}{36183 (S)} & 13596 & Battlefield 1 & Electronic Arts & PlayStation 4 & Games \\
          & 1257  & Dualshock 2 Analog Wired Controller SCPH-10010 & Retro Gaming \& Microconsoles & PlayStation 2 & Accessories \\
          & 9470  & SimCity [Instant Access] & Electronic Arts & PC    & Games \\
          & 8491  & SimCity: Limited Editio & Electronic Arts & PC    & Games \\
          & \underline{789}   & Fallout 1 / Fallout 2 Bundle (Jewel Case) - PC & Interplay & PC    & Games \\
          \midrule
    \multirow{5}[1]{*}{7849 (F)} & 3488  & Wall Charger for Nintendo DS Lite & Chargers & Nintendo DS & Accessories \\
          & 1887  & Rollercoaster Tycoon 2: Triple Thrill Pack - PC & /     & PC    & Games \\
          & 2869  & Nintendo DS Lite Polar White & /     & Nintendo DS & Consoles \\
          & 6952  & Wii Remote Plus - Black & /     & Nintendo & Accessories \\
          & \underline{7550}  & Kinect Disneyland Adventures - Xbox 360 & Kids \& Family & Xbox 360 & Games \\
    \bottomrule
    \end{tabular}%
    }
  \label{tab:cases}%
\end{table*}%

\subsection{Discussion on Inference Costs} 
As shown in Table~\ref{tab:efficient}, the majority of our model's inference latency is attributed to the LLM’s reasoning phase, which accounts for approximately 88.60\% of the total inference time across all four datasets.
Nevertheless, in line with the “deep thinking” paradigm adopted by recent advanced large language models such as DeepSeek-R1 and GPT-5, we suggest that this "latency" is essential, as it provides transparency into how the model generates recommendations and thereby fosters user trust.
Given these considerations, we propose that the most suitable application scenarios for our model are in personalized conversational recommendation systems with user interfaces, rather than as a back-end algorithm for large-scale online recommendation tasks involving billions of web requests.

\begin{table}[t]
  \centering
    \captionsetup{font={small}}
  \caption{Average inference time (seconds) per batch.}
    \vskip -0.1in
  \scalebox{0.9}{
    \begin{tabular}{cccc|ccc}
    \toprule
    Dataset & LlaRA & TIGER & TokenRec & \ourname{} & LLM & Diffusion \\
    \midrule
    LastFM & 0.4255  & 0.3502  & 0.0240  & 1.6110  & 1.5244  & 0.0866  \\
    ML1M  & 0.7370  & 0.5990  & 0.0367  & 2.4264  & 2.0994  & 0.3271  \\
    Games & 0.5281  & 0.4316  & 0.0273  & 1.7338  & 1.4920  & 0.2418  \\
    Beauty & 0.5423  & 0.4388  & 0.0292  & 1.7643  & 1.5385  & 0.2258  \\
    \bottomrule
    \end{tabular}%
    }
  \label{tab:efficient}%
\end{table}%

\subsection{Analysis of Generalizability}
To assess the Generalizability of our approach and its LLM-based counterparts, we utilize the Beauty and Games datasets. In these datasets, we exclude the 5\% of users with the least interaction history from the training split, treating them as newly added users in the evaluation set. The results demonstrate that advanced LLM-based methods (i.e., TokenRec, TIGER, LLaRA, and \ourname{}) maintain strong performance in this scenario. This can be attributed to their ability to leverage textual information (e.g., title and category) as supplementary cues for inferring user preferences when interaction history is limited. Furthermore, our hybrid retrieval process encourages \ourname{} to predict items with relevant categories and brands, making it particularly effective in cases involving previously unseen users.

\begin{table}[t]
  \centering
      \captionsetup{font={small}}
  \caption{Generalizability of the representative generative recommendation models to unseen users (5\%).}
  \vskip -0.1in
  \scalebox{0.9}{
    \begin{tabular}{ccccc}
    \toprule
    \multirow{2}[2]{*}{\textbf{Model}} & \multicolumn{2}{c}{\textbf{Beauty}} & \multicolumn{2}{c}{\textbf{Games}} \\
          & \textbf{HR@10} & \textbf{NDCG@10} & \textbf{HR@10} & \textbf{NDCG@10} \\
    \midrule
    P5    & 0.0255  & 0.0156  & 0.0643  & 0.0368  \\
    CoLLM & 0.0263  & 0.0171  & 0.0663  & 0.0373  \\
    TokenRec & 0.0374  & 0.0197  & 0.0747  & 0.0414  \\
    TIGER & 0.0384  & 0.0195  & 0.0759  & 0.0429  \\
    LlaRA & 0.0380  & 0.0198  & 0.0770  & 0.0436  \\
    \ourname{}  & 0.0391  & 0.0205  & 0.0787  & 0.0450  \\
    \bottomrule
    \end{tabular}%
    }
  \label{tab:unseen}%
\end{table}%

\subsection{Implementation Details}
The proposed model is developed based on Hugging Face and PyTorch, leveraging multiple NVIDIA A800 GPUs (80GB).
We employ the \emph{meta-Llama/Llama-3.2.-1B-Instruct} model as the foundation of our LLM, which is subsequently quantized into a 4-bit format to facilitate efficient fine-tuning. In \ourname{}, the trainable parameters are divided into three parts, with their respective sizes detailed in Table~\ref{tab:param}. Additionally, the total model parameter size is estimated to be 3,058.967 MB.
To maximize the utilization of our computing hardware, the NVIDIA A800 GPU (80GB), we have configured a batch size of 24 and a maximum of 200 epochs. The training process on the four datasets (ML1M, Games, Beauty, and LastFM) demands considerable time, exceeding 72 hours, 60 hours, 50 hours, and 24 hours on a single A800 GPU, respectively.
For all baselines, we conduct negative sampling from the uniform distribution at the ratio of 1:1, which is not conducted in DreamRec and \ourname{}.

\begin{table}[t]
  \centering
      \captionsetup{font={small}}
  \caption{Statistics of the trainable parameters.}
  \vskip -0.1in
  \scalebox{1}{
    \begin{tabular}{cccc}
    \toprule
    \textbf{} & \textbf{Name} & \textbf{Type} & \textbf{Params} \\
    \midrule
    1     & model & PeftModelForCausalLM & 760 M \\
    2     & diffusion\_net & Transformer   & 5.8 M \\
    3     & item\_tokenizer & MLP & 2.6 M \\
    4     & user\_tokenizer & MLP & 2.6 M \\
    5 & condition\_net & MLP & 1.1 M \\
    \bottomrule
    \end{tabular}%
    }
  \label{tab:param}%
\end{table}%
\end{document}